\documentclass[fleqn,usenatbib]{mnras}

\usepackage{newtxtext,newtxmath}
\usepackage[T1]{fontenc}

\DeclareRobustCommand{\VAN}[3]{#2}
\let\VANthebibliography\thebibliography
\def\thebibliography{\DeclareRobustCommand{\VAN}[3]{##3}\VANthebibliography}

\usepackage{graphicx}	% Including figure files
\usepackage{amsmath}	% Advanced maths commands
\usepackage{physics}
\usepackage{subfigure}
\usepackage{chemformula}

\title[ML and fundamental constants via white dwarfs]{Refining fundamental constants with white dwarfs: machine learning informed constraints on fine-structure constant and proton-to-electron mass ratio}

\author[Uniyal et al.]{
Akhil Uniyal,$^{1}$\thanks{E-mail: akhil\_uniyal@sjtu.edu.cn}
Surajit Kalita,$^{2}$\thanks{E-mail: surajit.kalita@uct.ac.za; corresponding author} 
Yosuke Mizuno,$^{1,3,4}$\thanks{E-mail: mizuno@sjtu.edu.cn}
Sayan Chakrabarti,$^{5}$\thanks{E-mail: sayan.chakrabarti@iitg.ac.in} and
Yan Lu$^{6}$\thanks{E-mail: luyan@pjlab.org.cn}
\\
$^{1}$Tsung-Dao Lee Institute, Shanghai Jiao Tong University, 1 Lisuo Road, Shanghai 201210, People’s Republic of China\\
$^{2}$Astronomical Observatory, University of Warsaw, Al. Ujazdowskie 4, Warsaw 00478, Poland\\
$^{3}$School of Physics and Astronomy, Shanghai Jiao Tong University, 800 Dongchuan Road, Shanghai 200240, People’s Republic of China\\
$^{4}$Key Laboratory for Particle Astrophysics and Cosmology (MOE) and Shanghai Key Laboratory for Particle Physics and Cosmology,\\ Shanghai Jiao Tong University, 800 Dongchuan Road, Shanghai 200240, People's Republic of China\\
$^{5}$Department of Physics, Indian Institute of Technology, Guwahati 781039, India\\
$^{6}$Shanghai Artificial Intelligence Laboratory, Shanghai, China, Shanghai 201210, People’s Republic of China
}

\date{Accepted XXX. Received YYY; in original form ZZZ}

\pubyear{2023}

\begin{document}
\label{firstpage}
\pagerange{\pageref{firstpage}--\pageref{lastpage}}
\maketitle

% Abstract of the paper
\begin{abstract}
We explore the potential variation of two fundamental constants, the fine-structure constant $\alpha$ and the proton-to-electron mass ratio $\mu$, within the framework of modified gravity theories and finite-temperature effects. Utilising high-precision white dwarf observations from the {\it Gaia}-DR3 survey, we construct a robust mass–-radius relation using a Bayesian-inspired machine learning framework. This empirical relation is rigorously compared with theoretical predictions derived from scalar–tensor gravity models and temperature-dependent equations of state. Our results demonstrate that both underlying gravitational theory and temperature substantially influence the inferred constraints on $\alpha$ and $\mu$. We obtain the strongest constraints as $|\Delta\alpha/\alpha|=2.10^{+32.56}_{-39.26}\times10^{-7}$ and $|\Delta\mu/\mu|=1.61^{+37.16}_{-34.67}\times10^{-7}$ for modified gravity parameter $\gamma\simeq -3.69\times10^{13}\,\mathrm{cm}^2$, while for the finite temperature case, these are $|\Delta\alpha/\alpha|=1.60^{+37.31}_{-35.42}\times10^{-7}$ and $|\Delta\mu/\mu|=1.23^{+37.02}_{-35.71}\times10^{-7}$ for $T \simeq 1.1 \times 10^7\rm\, K$. These findings yield tighter constraints than those reported in earlier studies and underscore the critical roles of gravitational and thermal physics in testing the constancy of fundamental parameters.
\end{abstract}

% Select between one and six entries from the list of approved keywords.

%%%%%%%%%%%%%%%%%%%%%%%%%%%%%%%%%%%%%%%%%%%%%%%%%%

%%%%%%%%%%%%%%%%% BODY OF PAPER %%%%%%%%%%%%%%%%%%

\section{Introduction} \label{sec:intro}
Stars with initial masses below $(10\pm2)\rm\,M_\odot$ ultimately evolve into white dwarfs (WDs; \citealt{1996cost.book.....G,2018MNRAS.480.1547L}). Chandrasekhar demonstrated that a non‐rotating, non‐magnetised carbon–oxygen WD has an upper mass limit of approximately $1.44\rm\,M_\odot$, now known as the Chandrasekhar mass limit~\citep{1935MNRAS..95..207C}.  Exceeding this threshold triggers a thermonuclear explosion, observed as a Type Ia supernova (SN\,Ia). However, the existence of both over‐luminous and under‐luminous SNe\,Ia implies progenitors above and below the conventional limit, respectively~\citep{1992AJ....104.1543F,1997MNRAS.284..151M,2006Natur.443..308H,2010ApJ...713.1073S}.  While the Chandrasekhar mass limit remains a fundamental concept, its value may not be unique, and various modified gravity frameworks have been invoked to accommodate sub‐ and super‐Chandrasekhar mass WDs. Over the past decade, numerous theories, particularly those using different scalar-tensor gravity formalisms, have been developed to explain these mass limits in both relativistic and non-relativistic contexts~\citep{2018JCAP...09..007K,2021ApJ...909...65K,2022PhRvD.105b4028S,2022PhLB..82736942K,2022PhRvD.106l4010A}.

Similarly, in the existing literature, three primary equations of state (EoSs) are commonly used to describe the degenerate matter inside WDs: the Chandrasekhar EoS~\citep{1931ApJ....74...81C}, Salpeter EoS~\citep{1961ApJ...134..669S, 1969ApJ...155..183S}, and relativistic Feynman–Metropolis–Teller (RFMT) EoS~\citep{1949PhRv...75.1561F, 2011PhRvC..83d5805R}. Among these, the RFMT EoS is the most comprehensive, as it accounts for Coulomb interactions and local inhomogeneities in relativistic electron distributions, effectively generalising the Chandrasekhar and Salpeter forms. These additional corrections typically reduce the predicted WD masses while slightly increasing their radii compared to the results from simpler EoSs. A comparison of the strengths and limitations of the three EoSs was provided by~\cite{Rotondo:2011zz}. Frequently used polytropic EoSs represent limiting cases in the non-relativistic and relativistic regimes of the Chandrasekhar and Salpeter models~\citep{YaBZel'dovich_1966,1983bhwd.book.....S}. It is important to note that the RFMT EoS exhibits noticeable deviations from the other two only at relatively low densities, particularly below $10^4\,\mathrm{g\,cm^{-3}}$~\citep{deCarvalho:2013rea}, which are generally not realised in the cores of observed WDs, where the density is maximum and is responsible for the majority of the WD mass. Consequently, the impact of RFMT corrections on the global structure of observed WDs is minimal. Therefore, in this study, we adopt the Chandrasekhar EoS and consider its modification in the presence of a finite temperature.

On the other hand, fundamental constants play an important role in understanding physical systems. Among them, the proton-to-electron mass ratio $\mu$ and fine‐structure constant $\alpha$ provide crucial information regarding gravity and electromagnetism. Several cosmological observations constrained variations in $\mu$, yielding $\Delta\mu/\mu \lesssim10^{-5}$ over redshift range $2\le z\le3$ from H$_2$ absorption systems~\citep{king2011new, Bagdonaite:2011ab, Bagdonaite:2013eia, Bagdonaite:2014mfa, Dapra:2015yva, Ubachs:2015fro, Dapra:2016dqh, Le:2019ijj}. By comparing CO (7–6) and [CI] emission lines from a lensed galaxy at $z = 5.2$, \cite{Levshakov:2012kv} set early-Universe upper limit $\Delta\mu/\mu<2.0\times10^{-7}$ and $\Delta\alpha/\alpha<8.0\times10^{-6}$. A more stringent limit, $\Delta\mu/\mu=(0.0\pm1.0)\times10^{-7}$, was obtained from CH$_3$OH lines in the $z=0.89$ absorber toward PKS\,1830$-$211~\citep{Bagdonaite:2013sia}, although this method applies only at $z\le1$ when using NH$_3$ or CH$_3$OH transitions. Similarly, using 21-cm absorption spectra of four quasar, \cite{Rahmani:2012ze} obtained $\Delta\mu/\mu=(0.0\pm1.5)\times10^{-6}$ at $z_{\rm abs}\approx1.3$. At higher redshift of $z\approx3.17$ in the system J1337$+$3152, \cite{Srianand:2010un} showed the constraint relaxes to $\Delta\mu/\mu=(-1.7\pm1.7)\times10^{-6}$. Moreover, using the absorption spectra of quasar HS\,$1549+1919$, \cite{2014MNRAS.445..128E} reported $\Delta\alpha/\alpha = (-5.4\pm 3.3_\mathrm{stat}\pm 1.5_\mathrm{sys})\times10^{-6}$, while for HE\,$0515-4414$, \cite{Kotus:2016xxb} reported $\Delta\alpha/\alpha = (-1.42\pm 0.55_\mathrm{stat}\pm 0.65_\mathrm{sys})\times10^{-6}$. Morever, using [CI], [CII], and CO emission lines across a wide redshift range, \cite{Levshakov:2017ivg} found no evidence for variation in $\alpha$, placing limits of $\abs{\Delta\alpha/\alpha} < 3\times 10^{-7}$ locally, $\abs{\Delta\alpha/\alpha} < 4\times 10^{-7}$ in M33 galaxy, and $\abs{\Delta\alpha/\alpha} < 4\times 10^{-7}$ at $z = 5.7-6.4$. Additionally, from the absorption systems in two Subaru quasar (PG\,$0117+213$ and  HS\,$1946+7658$) spectra with weighted mean, \cite{Murphy:2017xaz} reported $\Delta\alpha/\alpha = (3.0\pm 2.8_\mathrm{stat}\pm 2.0_\mathrm{sys})\times10^{-6}$ and for PHL957, \cite{Webb:2024} reported $ \Delta\alpha/\alpha =-0.53^{+5.45}_{-5.51}\times10^{-6}$. These measurements are considered more robust because they are derived from high-resolution quasar absorption spectra, which are then compared with the quantum mechanical predictions of atomic and molecular transitions. The underlying cause of potential variations in these constants remains uncertain. One possibility is that it may reflect the surrounding environment, which is linked to cosmic evolution. Each quasar was observed at different redshifts corresponding to distinct cosmic epochs, and thus, each under its own conditions, such as ambient temperature or the influence of dark energy.

Over the years, several constraints on $\alpha$ and $\mu$ were also put from WD physics. Primarily, these analyses are performed in two ways. First, analysing the individual WD spectra, such as using H$_2$ lines in GD\,133 and G29$-$38, \cite{2014PhRvL.113l3002B} estimated $\Delta\mu/\mu=(-2.7\pm4.9)\times10^{-5}$ and $(-5.8\pm4.1)\times10^{-5}$, respectively. Later, using Fe\,V features in G191$-$B2B, \cite{2021MNRAS.500.1466H} calculated $\Delta\alpha/\alpha=(6.36\pm2.27)\times10^{-5}$. Another recent study from the same WD spectra has found $|\Delta\mu/\mu|=(-0.360\pm0.864)\times10^{-8}$~\citep{Le2021}. These methods rely on whether the fundamental constants vary in a strong gravitational potential found near WDs, as experienced by atoms and molecules in the relatively low-density photosphere~\citep{Khoury:2003aq}. This is related to the fact that fundamental constants appear as coupling constants in higher-dimensional theory, which is dependent on the background environment~\citep{2003Ap&SS.283..627B}. Close to massive compact objects, scalar degrees of freedom that couple to the electromagnetic, gravitational, or fermionic sectors can produce spatial variations in the effective values of these constants. 

The second method involves analysing the WD mass–radius relation, as their equation of state is well understood. Simulations of 100 WD mass--radius pairs in the range $0.3{\rm\,M_\odot}<M<1.2\rm\,M_\odot$ suggested $\Delta\alpha/\alpha=(2.7\pm9.1)\times10^{-5}$~\citep{2017PhRvD..96h3012M}; however, because they are not based on real observations, they likely underestimate the true uncertainties. Previously, using massive WDs from different {\it Gaia} datasets, we derived constraints on $\alpha$ and $\mu$ under both Newtonian and modified‐gravity frameworks, demonstrating that the choice of gravitational theory significantly influences these limits~\citep{Kalita:2023hcl}. This analysis confirmed that the underlying theory of gravity plays a crucial role in dense stellar environments and yields notably tighter bounds on the fundamental constants when evaluated within alternative gravity models. Subsequently, we extended our investigation to include the effects of WD temperature on the constraints of fundamental constants~\citep{Uniyal:2023bff}, providing the one of the tightest limits of $|\Delta\alpha/\alpha|=4.20\times10^{-7}$ and $|\Delta\mu/\mu|=3.23\times10^{-7}$. Apart from quasar and WD, more recently, using data from extragalactic localised fast radio bursts, \cite{2024MNRAS.533L..57K} and \cite{2025JCAP...01..059L} have placed some of the most stringent constraints on these parameters. Additional bounds from other astronomical and cosmological datasets are mentioned in \cite{Levshakov:2013oja,2015PhRvC..92a4319D,2015Ap&SS.357....4K,2017RPPh...80l6902M,2017PhRvC..96d5802M,2018MNRAS.474.1850H}.

In this study, to ensure more robust results, we use a machine learning (ML) methodology to empirically map the WD mass--radius relation by training a feedforward neural network on high-precision {\it Gaia}-DR3 data~\citep{2023MNRAS.518.5106J}. The dataset is split into training, validation, and test sets with standardised inputs and outputs. We incorporate observational uncertainties through custom-weighted loss and apply Monte Carlo (MC) dropout~\citep{gal2016dropout} during training and inference for uncertainty estimation. Therefore, we generated MC ensembles to extract the mean trends and confidence intervals, enabling precise estimates of the stellar parameters and rigorous constraints on $\alpha$ and $\mu$. 

The remainder of this paper is organised as follows. In Section~\ref{sec:ml}, we present the ML methodology used to derive the empirical WD mass--radius relation. Section~\ref{result} details the constraints on the fundamental constants obtained by incorporating the effects of modified gravity and finite temperature, using the ML curve as a reference. Finally, Section~\ref{sec:discussion} discusses our findings and concludes the study in Section~\ref{sec:conclusion}.

%%%%%%%%%%%%%%%%%%%%%%%%%%%%%%%%%%%%%%%%%%%%%%%%%%%%%%%%%%%%%%%%%%%%%%%%%%%%%%%%%%%%%%%%%%%%%%%%%%%%%%%%%%%%%%%%%%%%%%%

\section{Machine learning model for deriving white dwarf mass--radius relation} \label{sec:ml}
\begin{figure}
    \centering
    \includegraphics[scale=0.15]{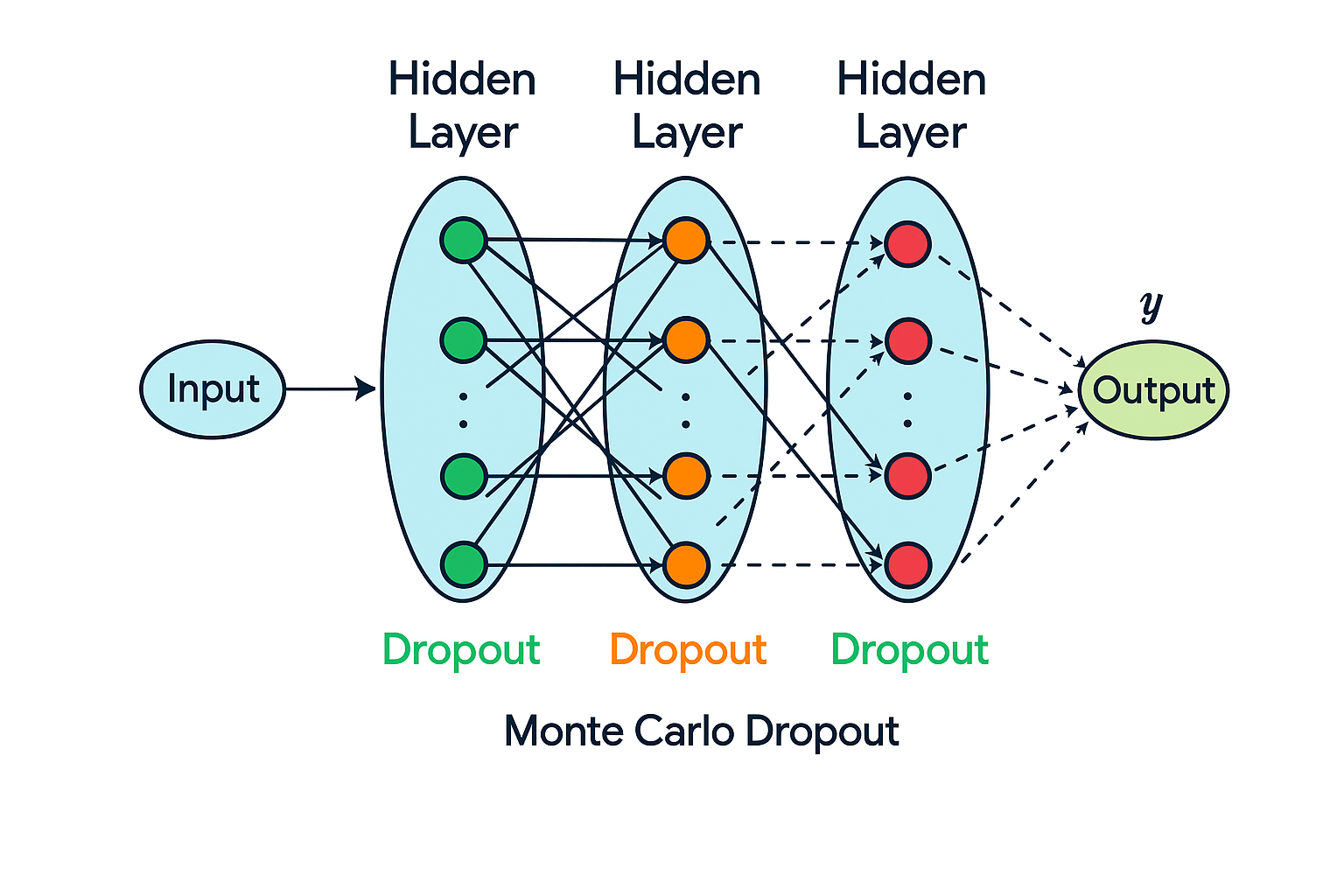}
    \caption{Schematic of the MC Dropout–enhanced feedforward neural network.}
    \label{nn}
\end{figure}

In this study, we develop and implement a fully data-driven computational framework to model the empirical relationship between WD radius and mass. This framework is based on a feedforward neural network architecture enhanced with MC Dropout for uncertainty quantification. Fig.~\ref{nn} illustrates the architecture of the proposed feedforward neural network. The model receives the normalised WD radii as inputs, which are propagated through multiple densely connected hidden layers. Dropout is applied during both training and inference stages at the hidden layers, thereby enabling stochastic forward passes that simulate MC sampling and facilitate the estimation of predictive uncertainties. The final output node yields the predicted stellar mass, whereas repeated inferences with active dropout produce a distribution of predictions, from which the mean and credible intervals are computed.

We begin our analysis by loading high-precision measurements of 8031 WDs within 100\,pc distance from the {\it Gaia}-DR3 catalogue~\citep{2023MNRAS.518.5106J}, containing only those candidates which are confirmed to be WDs with more than 90\% accuracy. For these selected objects, we employ their spectroscopically inferred masses alongside surface gravities ($\log g$) to calculate the stellar radii. The dataset is then partitioned into training, validation, and test subsets to facilitate robust model generalisation. All the input features and target values are standardised to zero mean and unit variance using statistics derived from the training set. Observational uncertainties are incorporated into the learning process through a custom loss function formulated as a modified weighted mean square loss criterion, enabling principled error propagation. Furthermore, we know that the final masses and radii of WDs depend on the evolutionary track of main-sequence stars. However, the WDs observed with \textit{Gaia} are the final stages of main-sequence stars. Hence, to understand the structure of WDs and the mass--radius relation, we require only hydrostatic balance equations along with an equation of state~\citep{1935MNRAS..95..207C}, and not the complete evolutionary history of stars.
\begin{figure}
    \centering
    \includegraphics[scale=0.33]{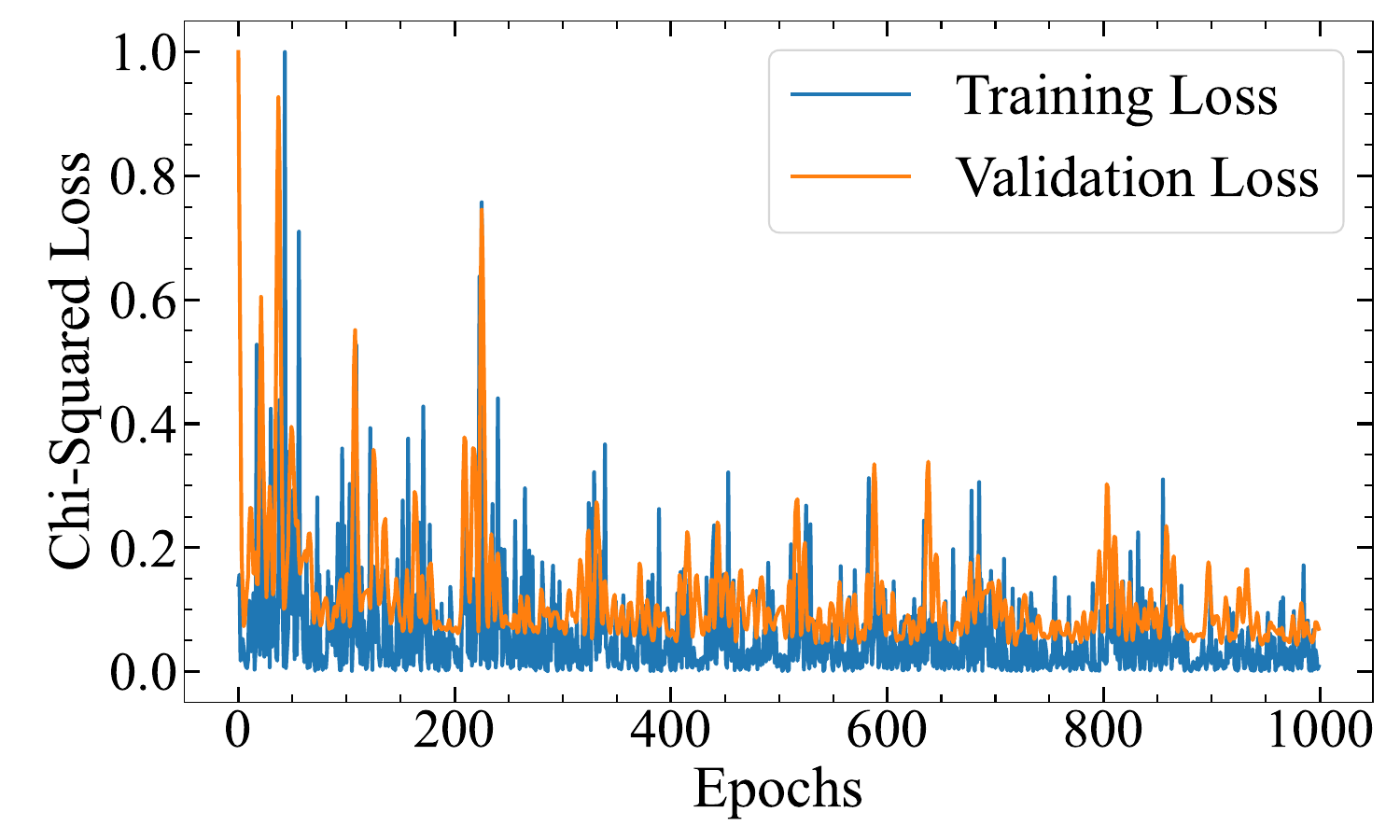}
    \caption{Training and Validation loss for the ML network.}
    \label{fig: chi}
\end{figure}
Fig.~\ref{fig: chi} shows the evolution of training and validation losses over successive epochs. The monotonic decrease in the training loss and a similar trend in the validation loss with slight fluctuations reflect effective learning and avoidance of overfitting. The slight discrepancy between the training and validation losses indicates appropriate regularisation. During each epoch, we introduce stochastic perturbations to the target variables, consistent with their associated observational uncertainties. This process acts as an additional regularisation mechanism and realistically reflects the measurement noise. After training the model for 1000 epochs using the Adam optimisation algorithm~\citep{kingma2014adam} and maintaining active dropout layers throughout, we perform multiple stochastic forward passes to generate the MC ensembles of model outputs. These ensembles are used to estimate the posterior mean and credible intervals for mass predictions, both for the test set and over a continuous grid of input radii. This approach seamlessly integrates error-aware learning with Bayesian-inspired uncertainty estimation, providing not only point predictions of stellar parameters, but also principled confidence bounds suitable for downstream astrophysical inference.

\begin{figure}
    \centering
    \subfigure[Green points represent {\it Gaia}-DR3 WDs within 100\,pc radius with corresponding error bars are shown in red.]{\includegraphics[scale=0.33]{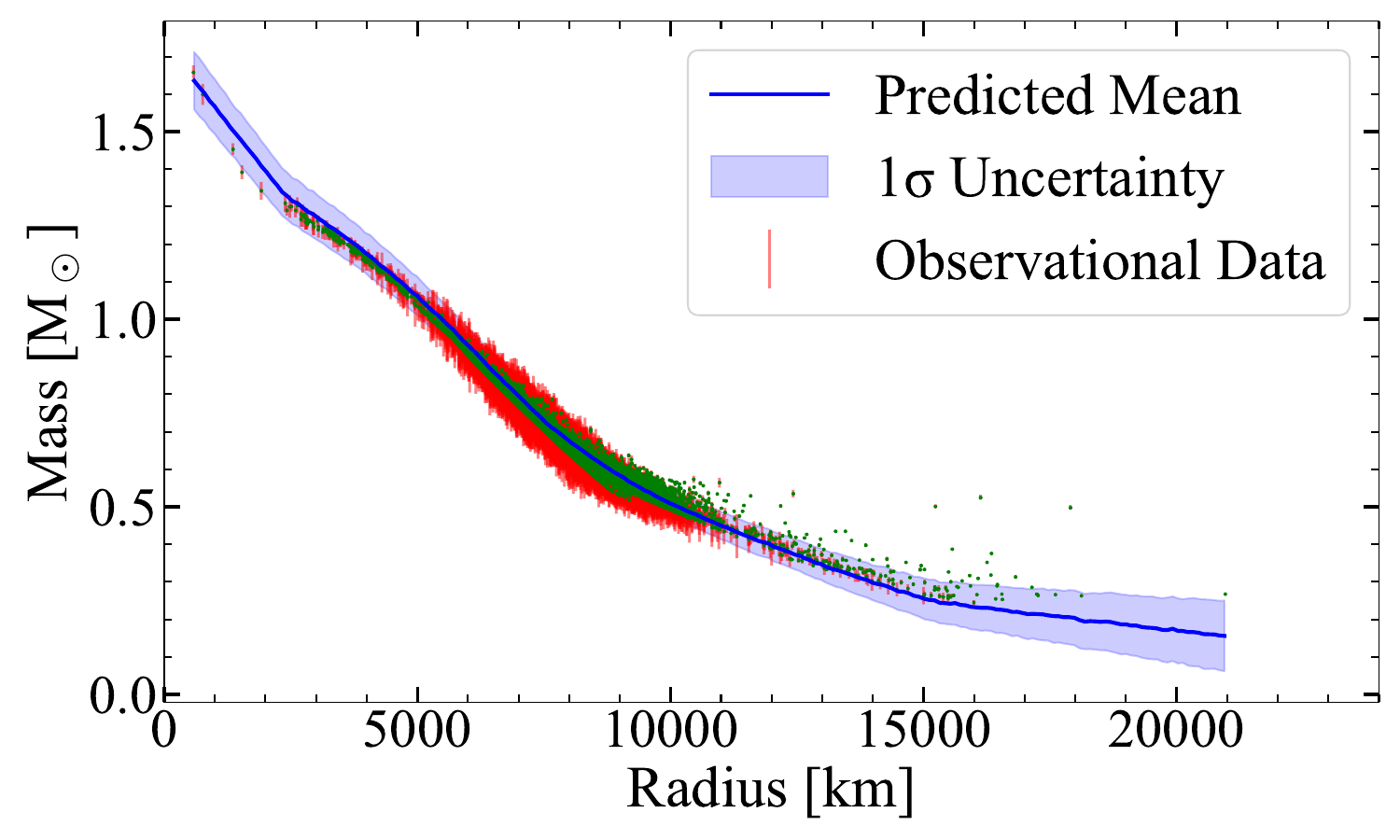}}    
    \subfigure[Theoretical curves showing the Chandrasekhar mass--radius curve along with its modifications due to finite temperature ($T\neq0 $) and modified gravity (modified gravity parameter $\gamma\neq0 $) effects.]{\includegraphics[scale=0.33]{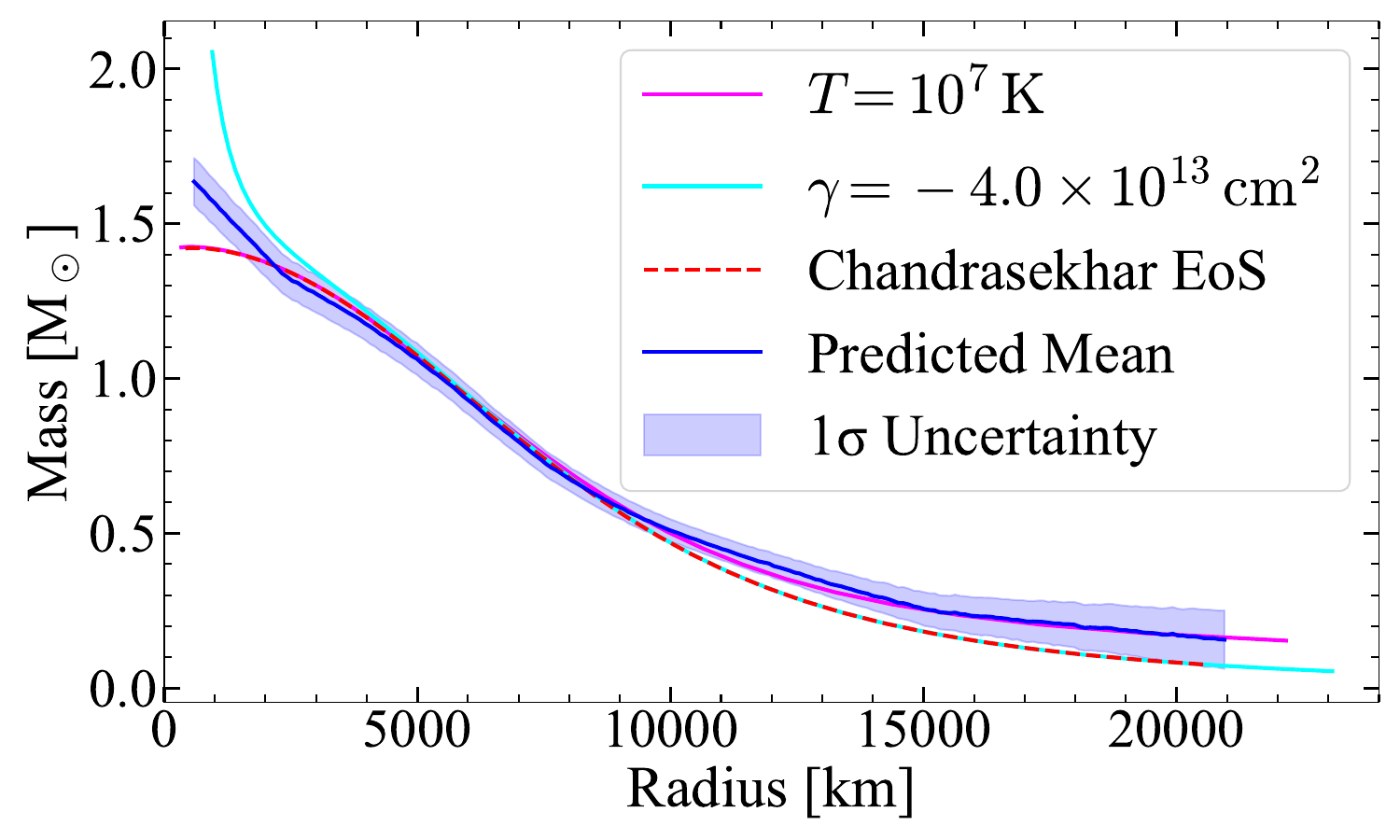}}    
    \caption{ML-predicted mass--radius relation of the WDs along with 1$\sigma$ uncertainty region derived from MC dropout.}
    \label{fig: fit}
\end{figure}

Fig.~\ref{fig: fit}(a) depicts the ML-predicted mass--radius curve along with the data points and their corresponding uncertainties. A comparison with the theoretical Chandrasekhar mass--radius relation is illustrated in Fig.~\ref{fig: fit}(b). The latter curve is derived by solving the equations of hydrostatic equilibrium
\begin{align}\label{pressure_wd}
    \dv{P(r)}{r} &= -\frac{Gm(r)\rho(r)}{r^2},\\
    \dv{m(r)}{r} &= 4\pi r^2 \rho(r),\label{mass_wd}
\end{align}
with the Chandrasekhar EoS~\citep{1935MNRAS..95..207C}
\begin{align}\label{e2.10}
    P &= \frac{\pi m_\text{e}^4 c^5}{3 h^3}\left[x_\text{F}\left(2x_\text{F}^2-3\right)\sqrt{x_\text{F}^2+1}+3\sinh^{-1}x_\text{F}\right],\nonumber\\
    \rho &= \frac{8\pi \mu_\text{e} m_\text{p}(m_\text{e}c)^3}{3h^3}x_\text{F}^3.
\end{align}
Here, $P$ is the pressure, $\rho$ is the density at radius $r$ and $m$ is the mass enclosed within this radius. $m_\text{e}$ is the mass of an electron, $m_\text{p}$ is the mass of a proton, $G$ is the Newton gravitational constant, $c$ is the speed of light, $h$ is the Planck constant,  $x_\text{F}=p_\text{F}/m_\text{e} c$ with $p_\text{F}$ being the Fermi momentum, and $\mu_\text{e}$ the mean molecular weight per electron.

It is evident that there are discrepancies in both low- and high-mass regimes. To explain the deviation in the low-mass region, we invoke finite temperature effects in the EoS, whereas to explain the same in the high-mass region, we introduce a scalar-tensor theory of gravity. Fig.~\ref{fig: fit}(b) illustrates the improved agreement between the observations and theoretical models incorporating these extensions, as well as the consistency of the ML-predicted curve within the bounds of theoretical expectations. Notably, the predicted values and their uncertainty intervals align closely with both the observed data and modified theoretical models, indicating that the neural network captures the underlying astrophysical trend with high fidelity. Given this consistency, we therefore combine both effects and utilise the ML curve to derive the constraints on $\alpha$ and $\mu$.

%%%%%%%%%%%%%%%%%%%%%%%%%%%%%%%%%%%%%%%%%%%%%%%%%%%%%%%%%%%%%%%%%%%%%%%%%%%%%%%%%%%%%%%%%%%%%%%%%%%%%%%%%%%%%%%%%%%%%%%

\section{Constraints on $\alpha$ and $\mu$} \label{result}

In a phenomenological approach to determine the constraints on $\alpha$ and $\mu$, we assume that the particle mass and quantum chromodynamics (QCD) scale can vary with the fixed Planck mass. This leads to the uncertainties in electron and proton masses as~\citep{Coc:2006sx}
\begin{align}\label{e2.11}
    \frac{\Delta m_\text{e}}{m_\text{e}} &= \frac{1}{2}\left(1+\mathsf{S}\right)\frac{\Delta \alpha}{\alpha},\\
    \frac{\Delta m_\text{p}}{m_\text{p}} &= \left[\frac{4}{5}\mathsf{R} +\frac{1}{5} \left(1+\mathsf{S}\right)\right]\frac{\Delta \alpha}{\alpha},\label{e2.12}
\end{align}
where $\mathsf{R}$ and $\mathsf{S}$ are dimensionless phenomenological parameters. Hence, uncertainty in $\mu$ can be written as
\begin{equation}\label{e2.13}
    \frac{\Delta \mu}{\mu} = \left[\frac{4}{5}\mathsf{R}-\frac{3}{10}\left(1+\mathsf{S}\right)\right]\frac{\Delta \alpha}{\alpha}.
\end{equation}
For different observations, $\mathsf{R}$ and $\mathsf{S}$ vary, such as data from the Wilkinson Microwave Anisotropy Probe (WMAP) suggest that $\mathsf{R} \approx 36$ and $\mathsf{S} \approx 160$~\citep{Coc:2006sx}, and a dilaton-type model yields $\mathsf{R} \approx 109$ and $\mathsf{S} \approx 0$~\citep{nakashima2010constraining}. For our calculations, we consider $\mathsf{R}=278\pm24$ and $\mathsf{S}=742\pm65$~\citep{2014MmSAI..85..113M} obtained from the astrophysical observations of the BL Lac object PKS\,$1413+135$.

To provide constraints on $\alpha$ and $\mu$ utilizing the ML-predicted curve, we define the joint likelihood function as
\begin{equation}\label{likelihood}
\mathcal{L} = \prod_{k=1}^{N} \frac{1}{\sqrt{2\pi}\,\sigma_k} \exp\left[-\frac{\left(M_\text{th}(r_k)-M_k\right)^2}{2\,\sigma_k^2} \right],
\end{equation}
where $M_k$ and $\sigma_k$ are the observed \textit{Gaia} WD mass and its uncertainty at radius \(r_k\), \(M_\text{th}(r_k)\) is the corresponding value from the ML-predicted curve, and $N$ is the total number of data points in the sample. This function needs to be maximised with respect to $\Delta \alpha / \alpha$ to obtain its best-fit value.

%%%%%%%%%%%%%%%%%%%%%%%%%%%%%%%%%%%%%%%%%%%%%%%%%%%%%%%%%%%%%%%%%%%%%%%%%%%%%%%%%%%%%%%%%%%%%%%%%%%%%%%%%%%%%%%%%%%%%%%

\subsection{Effects of modified gravity} \label{mg}
To understand the deviation at the high-mass end, especially to account for WDs with masses more than the Chandrasekhar mass limit (often called super-Chandrasekhar WDs), we consider the effect of modified gravity. Note that although super-Chandrasekhar WDs are also expected to exist under high magnetic fields or rotations, it is evident from Fig.~\ref{fig: fit}(b) that the massive WDs follow a well-defined path, unlike the lighter ones. Had strong magnetic fields or fast rotation been the reasons for these massive WDs to exceed the Chandrasekhar mass limit, each one of them would need nearly the same field strength or angular frequency, which is unlikely given how differently they are distributed across the sky and thus their environments. Moreover, for intermediate and massive WDs, the data lie nearly on the standard Chandrasekhar mass--radius curve, which implies that magnetism and rotation have negligible effects on their structures in that range. Modified gravity, on the other hand, can shift the mass--radius relation for massive WDs without forcing them to share identical magnetic fields or spins. Finally, the mass of a WD exceeds the Chandrasekhar limit, providing central magnetic fields exceeding the Schwinger limit ($4.4 \times 10^{13}$\,G), which might make it unstable owing to its high magnetic-to-gravitational energy ratio~\citep{Braithwaite2009}.

In the case of modified gravity, we consider generalised scalar-tensor gravity. It was already shown that for such scalar-tensor gravity in the Newtonian limit, the Poisson equation is modified as~\citep{2020PhRvD.101f4050T,2017JCAP...10..004B}
\begin{equation}\label{e2.6p}
    \nabla^2 \Phi \approx 4\pi G (\rho - 2\gamma \nabla^2 \rho),
\end{equation}
where $\gamma$ denotes the modified gravity parameter. This further can be used to write the hydrostatic balance equations as~\citep{2016PhRvL.116o1103J,2023PhRvD.107d4072K}
\begin{align}\label{mg_pm}
    \dv{P}{r} &= -\frac{Gm\rho}{r^2}+8\pi G\gamma \rho \dv{\rho}{r},\\
    \dv{m}{r} &= 4\pi r^2 \rho.\label{e2.9}
\end{align}

\begin{figure}
    \centering
    \subfigure[$\Delta\alpha/\alpha$]{\includegraphics[scale=0.33]{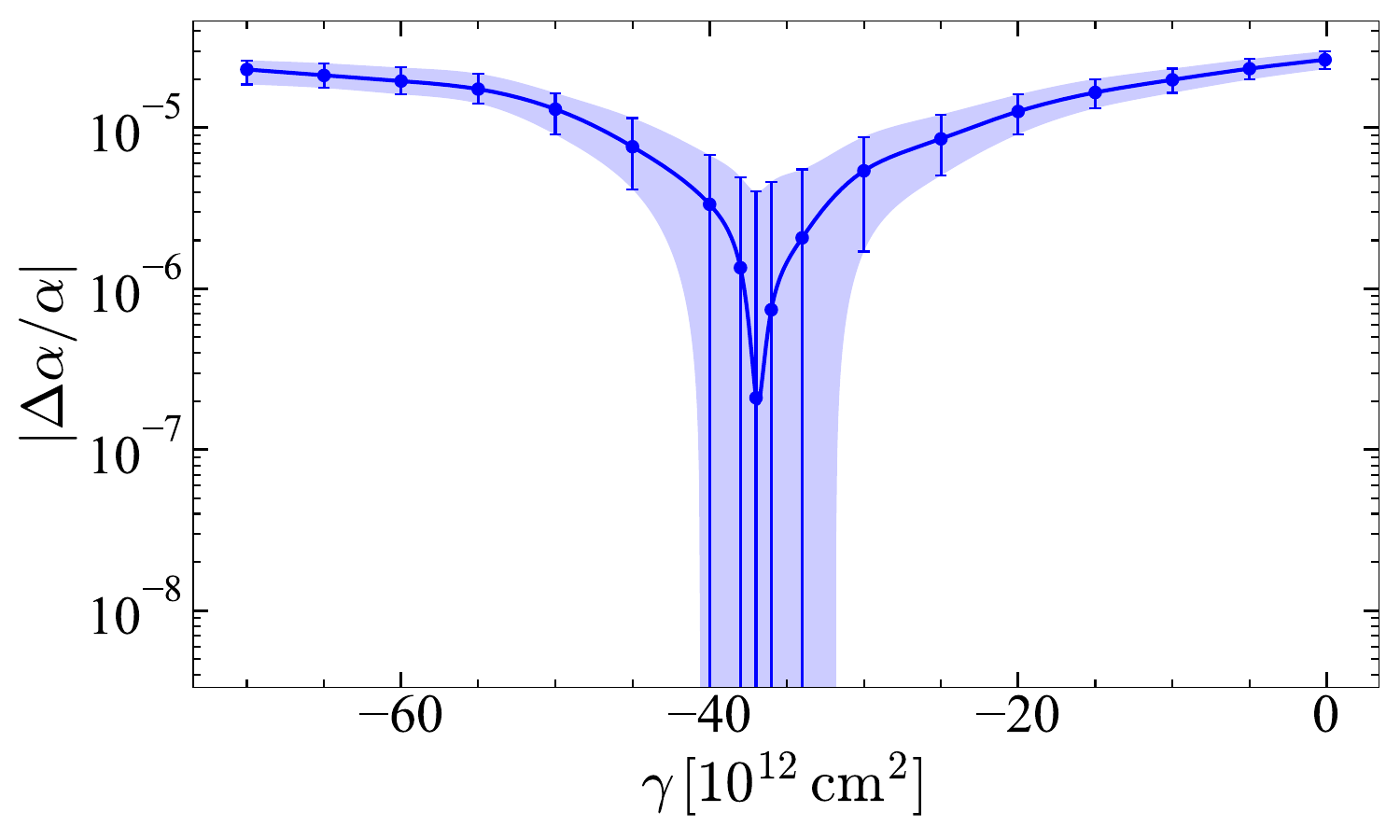}}    
    \subfigure[$\Delta\mu/\mu$]{\includegraphics[scale=0.33]{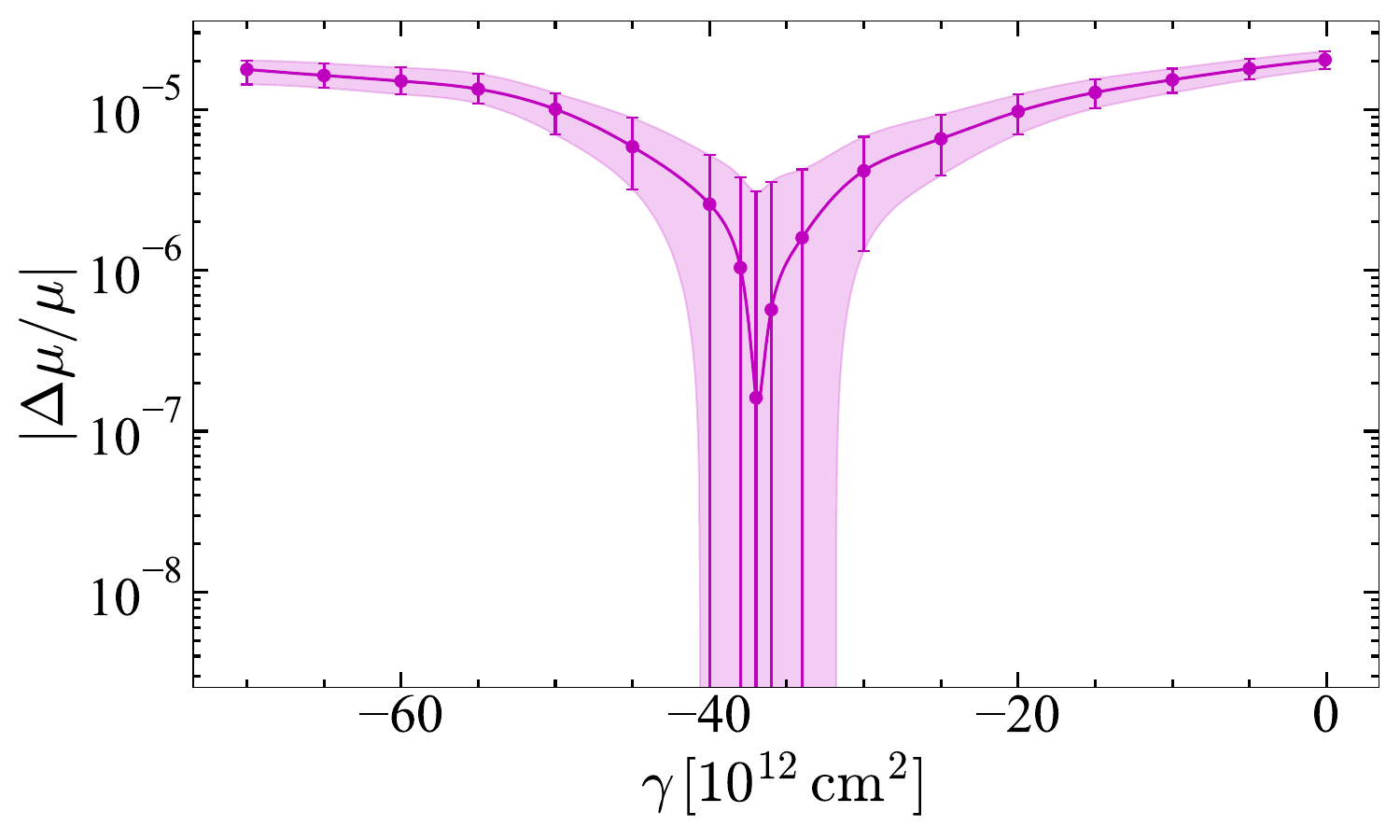}}    
    \caption{Variation of $\abs{\Delta\alpha/\alpha}$ and $\abs{\Delta\mu/\mu}$ along with 1$\sigma$ uncertainty region as a function of $\gamma$.}
    \label{fig: constants_mg}
\end{figure}

Therefore, we calculate mass--radius relations of this modified gravity inspired WDs for different $\gamma$ with the help of Chandrasekhar EoS. We compare these curves with the ML curve using the likelihood function of Eq.~\eqref{likelihood} and maximise it with respect to $\Delta \alpha / \alpha$. Fig.~\ref{fig: constants_mg} shows the maximised $\abs{\Delta\alpha/\alpha}$ and $\abs{\Delta\mu/\mu}$ values with 1$\sigma$ confidence intervals for different modified gravity parameter $\gamma$. Clearly, both $\abs{\Delta\alpha/\alpha}$ and $\abs{\Delta\mu/\mu}$ exhibit a decreasing trend until they reach a minimum, after which they increase indefinitely. This suggests that the gravitational model can affect the constraints on these fundamental constants. The best constraints are $\abs{\Delta\alpha/\alpha} = 2.10^{+32.56}_{-39.26}\times 10^{-7}$ and $\abs{\Delta\mu/\mu} = 1.61^{+37.16}_{-34.67}\times 10^{-7}$ for $\gamma \approx -3.69\times 10^{13}\rm\,cm^2$. These constraints are stronger than those in previous studies. Note that the ML algorithm, in principle, is able to generate as many missing data points as required and can be used to impose robust constraints on $\alpha$ and $\mu$.

%%%%%%%%%%%%%%%%%%%%%%%%%%%%%%%%%%%%%%%%%%%%%%%%%%%%%%%%%%%%%%%%%%%%%%%%%%%%%%%%%%%%%%%%%%%%%%%%%%%%%%%%%%%%%%%%%%%%%%%

\subsection{Effects of temperature} \label{temp}
For the temperature $T$, we follow a similar analysis as aforementioned case, however, in this case, the EoS is modified due to the presence of the temperature as~\citep{de2014relativistic},
\begin{align}\label{pe}
P &= \frac{16\pi \sqrt{2}}{h^3} m_\text{e}^4 c^5 \beta^{5/2} \left[ F_{3/2} (\eta,\beta)+ \frac{\beta}{2} F_{5/2} (\eta,\beta) \right],\\
\rho  &= \frac{8\pi \sqrt{2}}{h^3} \mu_\text{e} m_\text{p} m_\text{e}^3 c^3 \beta^{3/2} \left[F_{1/2}(\eta, \beta) + \beta F_{3/2}(\eta, \beta)\right],\label{rhoe}
\end{align}
where
\begin{equation}
F_{k} (\eta,\beta)=\int_{0} ^{\infty} \frac{t^k \sqrt{1+(\beta/2)t}}{1+ e^{t-\eta}}\dd{t}.
\label{eq:Fk1}
\bigskip
\end{equation} 
Here, $\beta = k_\text{B} T/m_\text{e} c^2$, where $k_\text{B}$ is the Boltzmann constant. The variable $t = \tilde{E}(p)/k_\text{B} T$ is a dimensionless parameter with $\tilde{E}(p) = \sqrt{p^2 c^2 + m_\text{e}^2 c^4} - m_\text{e} c^2$ representing the relativistic kinetic energy of an electron with momentum $p$. The parameter $\eta = \tilde{\mu}_\text{e}/k_\text{B} T$ denotes the chemical potential energy where $\tilde{\mu}_\text{e}$ is the electron chemical potential.

\begin{figure}
    \centering
    \subfigure[$\Delta\alpha/\alpha$]{\includegraphics[scale=0.33]{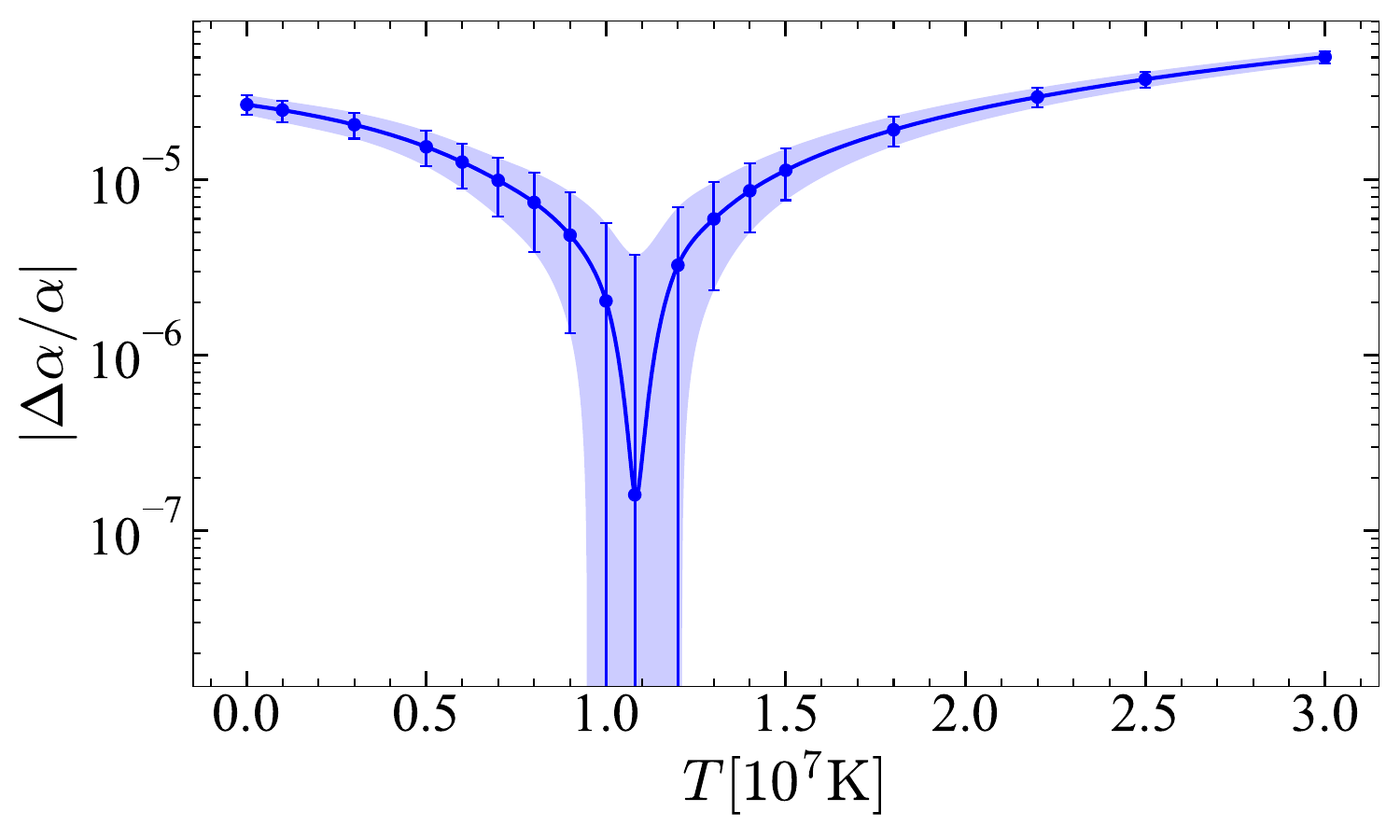}}    
    \subfigure[$\Delta\mu/\mu$]{\includegraphics[scale=0.33]{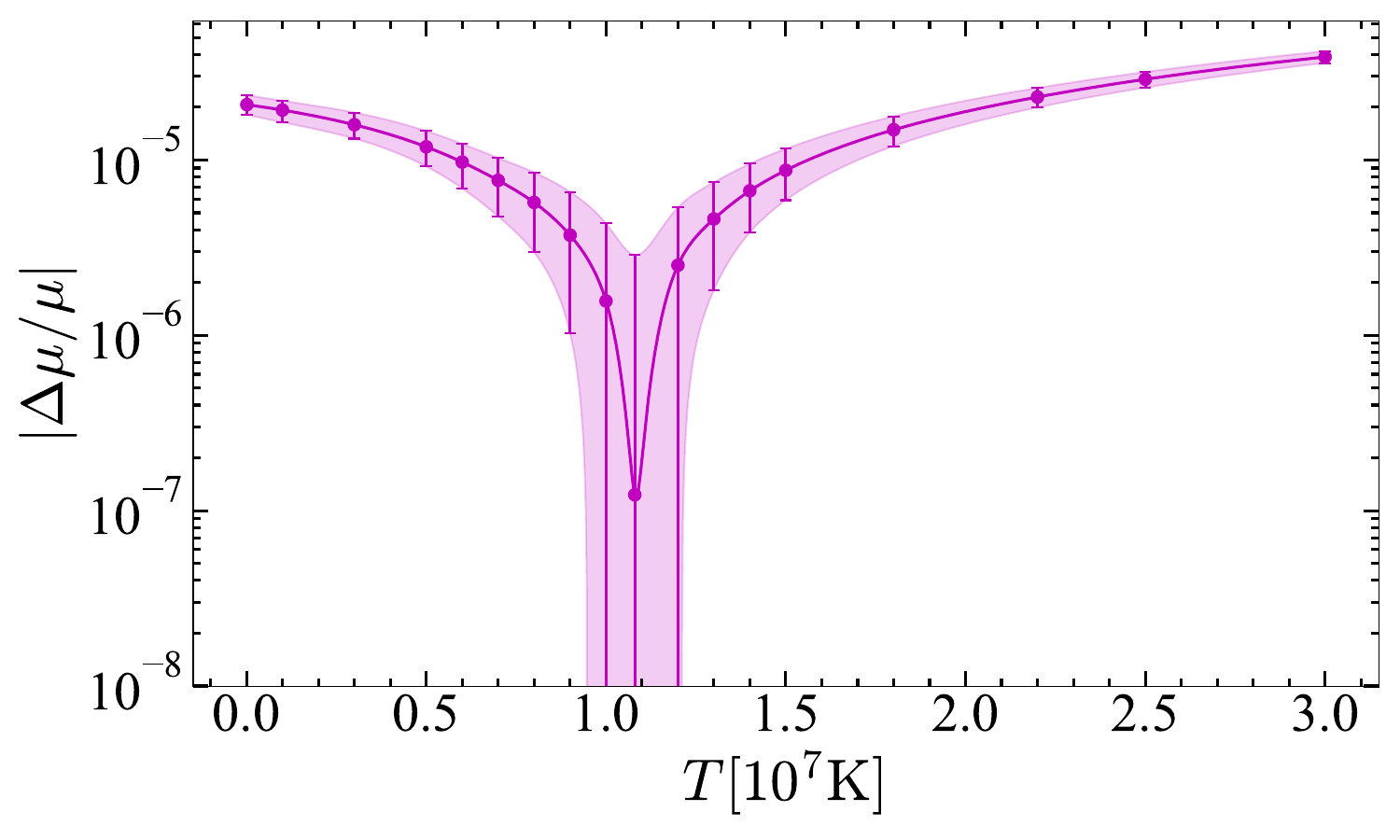}}    
    \caption{Same as Fig.~\ref{fig: constants_mg} but shown as a function of temperature.}
    \label{fig: constants_temp}
\end{figure}

Fig. \ref{fig: constants_temp} shows $\abs{\Delta\alpha/\alpha}$ and $\abs{\Delta\mu/\mu}$ with respect to $T$ with 1$\sigma$ confidence intervals, for which the likelihood function $\mathcal{L}$ is maximised. Similar to the case of modified gravity, they also show trends of first decreasing to minima and then continuing to increase with $T$. A further increase in $T$ causes these constraints to increase indefinitely, indicating that such a high temperature is not possible inside the WDs. The best constraints are $\abs{\Delta\alpha/\alpha} = 1.60^{+37.31}_{-35.42}\times 10^{-7}$ and $\abs{\Delta\mu/\mu} = 1.23^{+37.02}_{-35.71}\times 10^{-7}$ for $T=1.1 \times 10^7\rm\, K$.

%%%%%%%%%%%%%%%%%%%%%%%%%%%%%%%%%%%%%%%%%%%%%%%%%%%%%%%%%%%%%%%%%%%%%%%%%%%%%%%%%%%%%%%%%%%%%%%%%%%%%%%%%%%%%%%%%%%%%%%

\section{Discussion}\label{sec:discussion}
In astrophysical studies, the mass--radius relationship of astronomical objects presents a complex inverse problem characterised by sparse and noisy observational data. To address this challenge, we have employed a Bayesian neural network framework enhanced with MC Dropout for robust uncertainty quantification. Our methodology integrates measurement errors directly into the learning process through a customised weighted loss function and perturbative data augmentation during training, effectively propagating observational uncertainties into model predictions. The architecture utilises dropout layers activated during both the training and inference phases, thereby enabling probabilistic predictions through multiple stochastic forward passes. This approach not only captures the inherent non-linearities in the mass--radius relation but also provides statistically meaningful confidence intervals that are crucial for comparing empirical results with theoretical models. The implementation demonstrates how modern ML techniques can be adapted for astronomical inverse problems where traditional parametric fitting methods may be insufficient, particularly when handling heteroscedastic errors and requiring simultaneous model flexibility with uncertainty estimation.

To explain the deviations of the {\it Gaia}-DR3 observed masses and radii of WDs within 100\,pc radius, we have considered the modified Poisson equation of general scalar-tensor theory and solved the corresponding WD hydrostatic equilibrium equations for various values of the modified gravity parameter $\gamma$. The resulting WD mass--radius relations were then compared with our ML–predicted curve via the likelihood function, which was maximised over the relative variation of $\alpha$. Similarly, to explain the less massive WDs, we considered the temperature-dependent EoS and obtained different mass--radius curves, which were then compared with the ML-predicted curve utilising the likelihood function to maximise $\Delta\alpha/\alpha$. In both cases, the best-fit values of $|\Delta\alpha/\alpha|$ and $|\Delta\mu/\mu|$ decrease to a minimum before rising again as $\gamma$ increases, indicating that the choice of gravitational model and finite temperature effect influence the constraints on these fundamental constants. For modified gravity, we have obtained the best constraints as $|\Delta\alpha/\alpha|=2.10^{+32.56}_{-39.26}\times10^{-7}$ and $|\Delta\mu/\mu|=1.61^{+37.16}_{-34.67}\times10^{-7}$ for $\gamma\simeq -3.69\times10^{13}\,\mathrm{cm}^2$, while for the finite temperature case, these are $|\Delta\alpha/\alpha|=1.60^{+37.31}_{-35.42}\times10^{-7}$ and $|\Delta\mu/\mu|=1.23^{+37.02}_{-35.71}\times10^{-7}$ for $T \simeq 1.1 \times 10^7\rm\, K$. These values are significantly tighter than those of most of the previously reported constraints (while incorporating additional effects of gravitational and thermal physics). Even for the standard Chandrasekhar case with $T=0$ and $\gamma=0$, this study achieves an order-of-magnitude improvement in precision compared to our previous work in \cite{Kalita:2023hcl}. These results are statistically consistent with those reported by \cite{Uniyal:2023bff}. Any minor difference arise due to two reasons: 1. differences in the dataset used and 2. differences in the data analysis techniques applied. The order-of-magnitude improvement is attributable to the ability of the ML model to interpolate missing data points, thereby enhancing the statistical power of our analysis.

It is important to note that for small deviations from their terrestrial values, variations in fundamental constants can be expressed through a linear parametrisation~\citep{Damour:1994zq, Berengut:2013dta, Bainbridge:2017lsj}
\begin{equation}
\frac{\Delta \alpha}{\alpha} \simeq \zeta_{\alpha}\,\Delta\Phi,\qquad
\frac{\Delta (m_{\mathrm{p}}/m_{\mathrm{e}})}{(m_{\mathrm{p}}/m_{\mathrm{e}})} \simeq \zeta_{\mu}\,\Delta\Phi,\qquad
\frac{\Delta G}{G} \simeq \zeta_{G}\,\Delta\Phi,
\end{equation}
where $\Delta\Phi$ is the change in dimensionless gravitational potential relative to a laboratory reference. The coupling coefficients $\zeta_{i}$ quantify the strength of the interaction between the scalar field and each sector. For a spherical body of mass $M$ and radius $R$, the Newtonian gravitational potential is $\Phi={GM}/{Rc^{2}}$, so that, to leading order, the magnitude of local variation scales with the object's compactness $M/R$. WDs are compact objects with high $M/R$ providing a crucial astrophysical laboratory for testing such couplings. Any variation in $\alpha$ or $\mu$ modifies atomic and molecular energy levels affecting spectral line positions, fine-structure splitting, and relative transition strengths. Precision spectroscopy of WD atmospheres, combined with detailed atmospheric models and accurate corrections for gravitational redshift and pressure shifts, can place constraints on $\zeta_{i}$. These measurements require high resolution, high signal-to-noise spectra, and careful differential comparisons to laboratory wavelengths to isolate potential signatures of new physics from astrophysical or instrumental effects.

%%%%%%%%%%%%%%%%%%%%%%%%%%%%%%%%%%%%%%%%%%%%%%%%%%%%%%%%%%%%%%%%%%%%%%%%%%%%%%%%%%%%%%%%%%%%%%%%%%%%%%%%%%%%%%%%%%%%%%%

\section{Conclusion}\label{sec:conclusion}
Our study demonstrates that different observations yield different ranges of limits on $\alpha$ and $\mu$, depending on the dataset and technique employed, which inherently indicates their dependencies on different underlying system environments. Our work marks the first implementation of an ML approach that achieves substantially tighter constraints than most previously reported results. By training a neural network on high‐precision WD data, we have demonstrated that data‐driven interpolation can fill in otherwise inaccessible regions of the mass--radius plane, effectively `bridging gaps' in observational coverage. Consequently, our ML-informed curve not only produces existing theoretical and empirical relations but also shrinks the allowed window for fundamental constant variation. Variations in $\alpha$ and $\mu$ hold significant physical implications. $\alpha$ is a critical parameter for validating the precision of quantum electrodynamics (QED) and other theoretical frameworks, and any discrepancy in its measured value may indicate physics beyond the Standard Model. Similarly, in cosmology, $\mu$ is essential for analysing the spectra of distant galaxies and tracing the chemical evolution of the universe. Thus, variations in $\mu$ could suggest temporal or spatial changes in fundamental constants, offering insights into the dynamical processes of the early universe. In the future, expanding the training dataset, both in size and diversity, can reduce statistical noise and mitigate sampling biases in the learned mass--radius relation, thereby enabling even more stringent tests of fundamental constant variations.

%%%%%%%%%%%%%%%%%%%%%%%%%%%%%%%%%%%%%%%%%%%%%%%%%%%%%%%%%%%%%%%%%%%%%%%%%%%%%%%%%%%%%%%%%%%%%%%%%%%%%%%%%%%%%%%%%%%%%%%

\section*{Acknowledgments}
The authors would like to thank the anonymous reviewers for their constructive comments to improve the manuscript content. S.K. gratefully acknowledges the support and hospitality from ICTP, Italy during a research visit through the Associates Programme. This research is supported by the National Key R\&D Program of China (grant no. 2023YFE0101200), the National Natural Science Foundation of China (grant No.\,12273022), the Research Fund for Excellent International PhD Students (grant no. W2442004), and the Shanghai Municipality Orientation Program of Basic Research for International Scientists (grant no. 22JC1410600). S.K. is funded by the National Science Centre, Poland (grant no. 2023/49/B/ST9/02777). Simulations were performed on the TDLI-Astro cluster and Siyuan Mark-I at Shanghai Jiao Tong University.

%%%%%%%%%%%%%%%%%%%%%%%%%%%%%%%%%%%%%%%%%%%%%%%%%%%%%%%%%%%%%%%%%%%%%%%%%%%%%%%%%%%%%%%%%%%%%%%%%%%%%%%%%%%%%%%%%%%%%%%

\section*{Data availability}
The data underlying this article were accessed from \url{http://svocats.cab.inta-csic.es/wdw5}. The derived data generated in this research will be shared on reasonable request to the corresponding author.

%\appendix

%%%%%%%%%%%%%%%%%%%%%%%%%%%%%%%%%%%%%%%%%%%%%%%%%%%%%%%%%%%%%%%%%%%%%%%%%%%%%%%%%%%%%%%%%%%%%%%%%%%%%%%%%%%%%%%%%%%%%%%

%-------------------------------------------------------------------

\bibliographystyle{mnras}
\bibliography{bibliography} % if your bibtex file is called example.bib

\begin{thebibliography}{}
\makeatletter
\relax
\def\mn@urlcharsother{\let\do\@makeother \do\$\do\&\do\#\do\^\do\_\do\%\do\~}
\def\mn@doi{\begingroup\mn@urlcharsother \@ifnextchar [ {\mn@doi@} {\mn@doi@[]}}
\def\mn@doi@[#1]#2{\def\@tempa{#1}\ifx\@tempa\@empty \href {http://dx.doi.org/#2} {doi:#2}\else \href {http://dx.doi.org/#2} {#1}\fi \endgroup}
\def\mn@eprint#1#2{\mn@eprint@#1:#2::\@nil}
\def\mn@eprint@arXiv#1{\href {http://arxiv.org/abs/#1} {{\tt arXiv:#1}}}
\def\mn@eprint@dblp#1{\href {http://dblp.uni-trier.de/rec/bibtex/#1.xml} {dblp:#1}}
\def\mn@eprint@#1:#2:#3:#4\@nil{\def\@tempa {#1}\def\@tempb {#2}\def\@tempc {#3}\ifx \@tempc \@empty \let \@tempc \@tempb \let \@tempb \@tempa \fi \ifx \@tempb \@empty \def\@tempb {arXiv}\fi \@ifundefined {mn@eprint@\@tempb}{\@tempb:\@tempc}{\expandafter \expandafter \csname mn@eprint@\@tempb\endcsname \expandafter{\@tempc}}}

\bibitem[\protect\citeauthoryear{{Astashenok}, {Odintsov}  \& {Oikonomou}}{{Astashenok} et~al.}{2022}]{2022PhRvD.106l4010A}
{Astashenok} A.~V.,  {Odintsov} S.~D.,   {Oikonomou} V.~K.,  2022, \mn@doi [\prd] {10.1103/PhysRevD.106.124010}, \href {https://ui.adsabs.harvard.edu/abs/2022PhRvD.106l4010A} {106, 124010}

\bibitem[\protect\citeauthoryear{Bagdonaite, Murphy, Kaper  \& Ubachs}{Bagdonaite et~al.}{2012}]{Bagdonaite:2011ab}
Bagdonaite J.,  Murphy M.~T.,  Kaper L.,   Ubachs W.,  2012, \mn@doi [Mon. Not. Roy. Astron. Soc.] {10.1111/j.1365-2966.2011.20319.x}, 421, 419

\bibitem[\protect\citeauthoryear{Bagdonaite, Dapr\`a, Jansen, Bethlem, Ubachs, Muller, Henkel  \& Menten}{Bagdonaite et~al.}{2013}]{Bagdonaite:2013sia}
Bagdonaite J.,  Dapr\`a M.,  Jansen P.,  Bethlem H.~L.,  Ubachs W.,  Muller S.,  Henkel C.,   Menten K.~M.,  2013, \mn@doi [Phys. Rev. Lett.] {10.1103/PhysRevLett.111.231101}, 111, 231101

\bibitem[\protect\citeauthoryear{Bagdonaite, Salumbides, Preval, Barstow, Barrow, Murphy  \& Ubachs}{Bagdonaite et~al.}{2014a}]{Bagdonaite:2014mfa}
Bagdonaite J.,  Salumbides E.~J.,  Preval S.~P.,  Barstow M.~A.,  Barrow J.~D.,  Murphy M.~T.,   Ubachs W.,  2014a, \mn@doi [Phys. Rev. Lett.] {10.1103/PhysRevLett.113.123002}, 113, 123002

\bibitem[\protect\citeauthoryear{{Bagdonaite}, {Salumbides}, {Preval}, {Barstow}, {Barrow}, {Murphy}  \& {Ubachs}}{{Bagdonaite} et~al.}{2014b}]{2014PhRvL.113l3002B}
{Bagdonaite} J.,  {Salumbides} E.~J.,  {Preval} S.~P.,  {Barstow} M.~A.,  {Barrow} J.~D.,  {Murphy} M.~T.,   {Ubachs} W.,  2014b, \mn@doi [\prl] {10.1103/PhysRevLett.113.123002}, \href {https://ui.adsabs.harvard.edu/abs/2014PhRvL.113l3002B} {113, 123002}

\bibitem[\protect\citeauthoryear{Bagdonaite, Ubachs, Murphy  \& Whitmore}{Bagdonaite et~al.}{2014c}]{Bagdonaite:2013eia}
Bagdonaite J.,  Ubachs W.,  Murphy M.~T.,   Whitmore J.~B.,  2014c, \mn@doi [Astrophys. J.] {10.1088/0004-637X/782/1/10}, 782, 10

\bibitem[\protect\citeauthoryear{Bainbridge et~al.}{Bainbridge et~al.}{2017}]{Bainbridge:2017lsj}
Bainbridge M.~B.,  et~al., 2017, \mn@doi [Universe] {10.3390/universe3020032}, 3, 32

\bibitem[\protect\citeauthoryear{{Banerjee}, {Shankar}  \& {Singh}}{{Banerjee} et~al.}{2017}]{2017JCAP...10..004B}
{Banerjee} S.,  {Shankar} S.,   {Singh} T.~P.,  2017, \mn@doi [\jcap] {10.1088/1475-7516/2017/10/004}, \href {https://ui.adsabs.harvard.edu/abs/2017JCAP...10..004B} {2017, 004}

\bibitem[\protect\citeauthoryear{Berengut, Flambaum, Ong, Webb, Barrow, Barstow, Preval  \& Holberg}{Berengut et~al.}{2013}]{Berengut:2013dta}
Berengut J.~C.,  Flambaum V.~V.,  Ong A.,  Webb J.~K.,  Barrow J.~D.,  Barstow M.~A.,  Preval S.~P.,   Holberg J.~B.,  2013, \mn@doi [Phys. Rev. Lett.] {10.1103/PhysRevLett.111.010801}, 111, 010801

\bibitem[\protect\citeauthoryear{Braithwaite}{Braithwaite}{2009}]{Braithwaite2009}
Braithwaite J.,  2009, \mn@doi [Monthly Notices of the Royal Astronomical Society] {10.1111/j.1365-2966.2008.14034.x}, 397, 763

\bibitem[\protect\citeauthoryear{{Brax}, {van de Bruck}, {Davis}  \& {Rhodes}}{{Brax} et~al.}{2003}]{2003Ap&SS.283..627B}
{Brax} P.,  {van de Bruck} C.,  {Davis} A.~C.,   {Rhodes} C.~S.,  2003, \mn@doi [\apss] {10.1023/A:1022543206870}, \href {https://ui.adsabs.harvard.edu/abs/2003Ap&SS.283..627B} {283, 627}

\bibitem[\protect\citeauthoryear{{Chandrasekhar}}{{Chandrasekhar}}{1931}]{1931ApJ....74...81C}
{Chandrasekhar} S.,  1931, \mn@doi [\apj] {10.1086/143324}, \href {https://ui.adsabs.harvard.edu/abs/1931ApJ....74...81C} {74, 81}

\bibitem[\protect\citeauthoryear{{Chandrasekhar}}{{Chandrasekhar}}{1935}]{1935MNRAS..95..207C}
{Chandrasekhar} S.,  1935, \mn@doi [\mnras] {10.1093/mnras/95.3.207}, \href {http://adsabs.harvard.edu/abs/1935MNRAS..95..207C} {95, 207}

\bibitem[\protect\citeauthoryear{Coc, Nunes, Olive, Uzan  \& Vangioni}{Coc et~al.}{2007}]{Coc:2006sx}
Coc A.,  Nunes N.~J.,  Olive K.~A.,  Uzan J.-P.,   Vangioni E.,  2007, \mn@doi [Phys. Rev. D] {10.1103/PhysRevD.76.023511}, 76, 023511

\bibitem[\protect\citeauthoryear{Damour \& Polyakov}{Damour \& Polyakov}{1994}]{Damour:1994zq}
Damour T.,  Polyakov A.~M.,  1994, \mn@doi [Nucl. Phys. B] {10.1016/0550-3213(94)90143-0}, 423, 532

\bibitem[\protect\citeauthoryear{Dapr{\`a}, Bagdonaite, Murphy  \& Ubachs}{Dapr{\`a} et~al.}{2015}]{Dapra:2015yva}
Dapr{\`a} M.,  Bagdonaite J.,  Murphy M.~T.,   Ubachs W.,  2015, \mn@doi [Mon. Not. Roy. Astron. Soc.] {10.1093/mnras/stv1998}, 454, 489

\bibitem[\protect\citeauthoryear{Dapr{\`a}, van~der Laan, Murphy  \& Ubachs}{Dapr{\`a} et~al.}{2017}]{Dapra:2016dqh}
Dapr{\`a} M.,  van~der Laan M.,  Murphy M.~T.,   Ubachs W.,  2017, \mn@doi [Mon. Not. Roy. Astron. Soc.] {10.1093/mnras/stw3003}, 465, 4057

\bibitem[\protect\citeauthoryear{{Davis} \& {Hamdan}}{{Davis} \& {Hamdan}}{2015}]{2015PhRvC..92a4319D}
{Davis} E.~D.,  {Hamdan} L.,  2015, \mn@doi [\prc] {10.1103/PhysRevC.92.014319}, \href {https://ui.adsabs.harvard.edu/abs/2015PhRvC..92a4319D} {92, 014319}

\bibitem[\protect\citeauthoryear{{Evans} et~al.,}{{Evans} et~al.}{2014}]{2014MNRAS.445..128E}
{Evans} T.~M.,  et~al., 2014, \mn@doi [\mnras] {10.1093/mnras/stu1754}, \href {https://ui.adsabs.harvard.edu/abs/2014MNRAS.445..128E} {445, 128}

\bibitem[\protect\citeauthoryear{{Feynman}, {Metropolis}  \& {Teller}}{{Feynman} et~al.}{1949}]{1949PhRv...75.1561F}
{Feynman} R.~P.,  {Metropolis} N.,   {Teller} E.,  1949, \mn@doi [Physical Review] {10.1103/PhysRev.75.1561}, \href {https://ui.adsabs.harvard.edu/abs/1949PhRv...75.1561F} {75, 1561}

\bibitem[\protect\citeauthoryear{{Filippenko} et~al.,}{{Filippenko} et~al.}{1992}]{1992AJ....104.1543F}
{Filippenko} A.~V.,  et~al., 1992, \mn@doi [\aj] {10.1086/116339}, \href {https://ui.adsabs.harvard.edu/abs/1992AJ....104.1543F} {104, 1543}

\bibitem[\protect\citeauthoryear{Gal \& Ghahramani}{Gal \& Ghahramani}{2016}]{gal2016dropout}
Gal Y.,  Ghahramani Z.,  2016, in Balcan M.~F.,  Weinberger K.~Q.,  eds,  Proceedings of Machine Learning Research Vol. 48, Proceedings of The 33rd International Conference on Machine Learning. PMLR, New York, New York, USA, pp 1050--1059 (\mn@eprint {arXiv} {2410.00684}), \url {https://proceedings.mlr.press/v48/gal16.html}

\bibitem[\protect\citeauthoryear{{Glendenning}}{{Glendenning}}{1996}]{1996cost.book.....G}
{Glendenning} N.~K.,  1996, {Compact Stars}.
Springer New York, NY, \mn@doi{10.1007/978-1-4684-0491-3}

\bibitem[\protect\citeauthoryear{{Hart} \& {Chluba}}{{Hart} \& {Chluba}}{2018}]{2018MNRAS.474.1850H}
{Hart} L.,  {Chluba} J.,  2018, \mn@doi [\mnras] {10.1093/mnras/stx2783}, \href {https://ui.adsabs.harvard.edu/abs/2018MNRAS.474.1850H} {474, 1850}

\bibitem[\protect\citeauthoryear{{Howell} et~al.,}{{Howell} et~al.}{2006}]{2006Natur.443..308H}
{Howell} D.~A.,  et~al., 2006, \mn@doi [\nat] {10.1038/nature05103}, \href {http://adsabs.harvard.edu/abs/2006Natur.443..308H} {443, 308}

\bibitem[\protect\citeauthoryear{{Hu} et~al.,}{{Hu} et~al.}{2021}]{2021MNRAS.500.1466H}
{Hu} J.,  et~al., 2021, \mn@doi [\mnras] {10.1093/mnras/staa3066}, \href {https://ui.adsabs.harvard.edu/abs/2021MNRAS.500.1466H} {500, 1466}

\bibitem[\protect\citeauthoryear{{Jain}, {Kouvaris}  \& {Nielsen}}{{Jain} et~al.}{2016}]{2016PhRvL.116o1103J}
{Jain} R.~K.,  {Kouvaris} C.,   {Nielsen} N.~G.,  2016, \mn@doi [\prl] {10.1103/PhysRevLett.116.151103}, \href {https://ui.adsabs.harvard.edu/abs/2016PhRvL.116o1103J} {116, 151103}

\bibitem[\protect\citeauthoryear{{Jim{\'e}nez-Esteban}, {Torres}, {Rebassa-Mansergas}, {Cruz}, {Murillo-Ojeda}, {Solano}, {Rodrigo}  \& {Camisassa}}{{Jim{\'e}nez-Esteban} et~al.}{2023}]{2023MNRAS.518.5106J}
{Jim{\'e}nez-Esteban} F.~M.,  {Torres} S.,  {Rebassa-Mansergas} A.,  {Cruz} P.,  {Murillo-Ojeda} R.,  {Solano} E.,  {Rodrigo} C.,   {Camisassa} M.~E.,  2023, \mn@doi [\mnras] {10.1093/mnras/stac3382}, \href {https://ui.adsabs.harvard.edu/abs/2023MNRAS.518.5106J} {518, 5106}

\bibitem[\protect\citeauthoryear{{Kalita}}{{Kalita}}{2024}]{2024MNRAS.533L..57K}
{Kalita} S.,  2024, \mn@doi [\mnras] {10.1093/mnrasl/slae062}, \href {https://ui.adsabs.harvard.edu/abs/2024MNRAS.533L..57K} {533, L57}

\bibitem[\protect\citeauthoryear{{Kalita} \& {Mukhopadhyay}}{{Kalita} \& {Mukhopadhyay}}{2018}]{2018JCAP...09..007K}
{Kalita} S.,  {Mukhopadhyay} B.,  2018, \mn@doi [\jcap] {10.1088/1475-7516/2018/09/007}, \href {http://adsabs.harvard.edu/abs/2018JCAP...09..007K} {9, 007}

\bibitem[\protect\citeauthoryear{{Kalita} \& {Mukhopadhyay}}{{Kalita} \& {Mukhopadhyay}}{2021}]{2021ApJ...909...65K}
{Kalita} S.,  {Mukhopadhyay} B.,  2021, \mn@doi [\apj] {10.3847/1538-4357/abddb8}, \href {https://ui.adsabs.harvard.edu/abs/2021ApJ...909...65K} {909, 65}

\bibitem[\protect\citeauthoryear{{Kalita} \& {Sarmah}}{{Kalita} \& {Sarmah}}{2022}]{2022PhLB..82736942K}
{Kalita} S.,  {Sarmah} L.,  2022, \mn@doi [Physics Letters B] {10.1016/j.physletb.2022.136942}, \href {https://ui.adsabs.harvard.edu/abs/2022PhLB..82736942K} {827, 136942}

\bibitem[\protect\citeauthoryear{Kalita \& Uniyal}{Kalita \& Uniyal}{2023}]{Kalita:2023hcl}
Kalita S.,  Uniyal A.,  2023, \mn@doi [Astrophys. J.] {10.3847/1538-4357/accf1c}, 949, 62

\bibitem[\protect\citeauthoryear{{Kalita}, {Sarmah}  \& {Wojnar}}{{Kalita} et~al.}{2023}]{2023PhRvD.107d4072K}
{Kalita} S.,  {Sarmah} L.,   {Wojnar} A.,  2023, \mn@doi [\prd] {10.1103/PhysRevD.107.044072}, \href {https://ui.adsabs.harvard.edu/abs/2023PhRvD.107d4072K} {107, 044072}

\bibitem[\protect\citeauthoryear{Khoury \& Weltman}{Khoury \& Weltman}{2004}]{Khoury:2003aq}
Khoury J.,  Weltman A.,  2004, \mn@doi [Phys. Rev. Lett.] {10.1103/PhysRevLett.93.171104}, 93, 171104

\bibitem[\protect\citeauthoryear{{King}, {Murphy}, {Ubachs}  \& {Webb}}{{King} et~al.}{2011}]{king2011new}
{King} J.~A.,  {Murphy} M.~T.,  {Ubachs} W.,   {Webb} J.~K.,  2011, \mn@doi [\mnras] {10.1111/j.1365-2966.2011.19460.x}, \href {https://ui.adsabs.harvard.edu/abs/2011MNRAS.417.3010K} {417, 3010}

\bibitem[\protect\citeauthoryear{{Kingma} \& {Ba}}{{Kingma} \& {Ba}}{2015}]{kingma2014adam}
{Kingma} D.~P.,  {Ba} J.,  2015, in Bengio Y.,  LeCun Y.,  eds, 3rd International Conference on Learning Representations, {ICLR} 2015, San Diego, CA, USA, May 7-9, 2015, Conference Track Proceedings.  (\mn@eprint {arXiv} {1412.6980}), \url {http://arxiv.org/abs/1412.6980}

\bibitem[\protect\citeauthoryear{Kotu{\v{s}}, Murphy  \& Carswell}{Kotu{\v{s}} et~al.}{2017}]{Kotus:2016xxb}
Kotu{\v{s}} S.~M.,  Murphy M.~T.,   Carswell R.~F.,  2017, \mn@doi [Mon. Not. Roy. Astron. Soc.] {10.1093/mnras/stw2543}, 464, 3679

\bibitem[\protect\citeauthoryear{{Kraiselburd}, {Landau}, {Negrelli}  \& {Garc{\'\i}a-Berro}}{{Kraiselburd} et~al.}{2015}]{2015Ap&SS.357....4K}
{Kraiselburd} L.,  {Landau} S.~J.,  {Negrelli} C.,   {Garc{\'\i}a-Berro} E.,  2015, \mn@doi [\apss] {10.1007/s10509-015-2325-4}, \href {https://ui.adsabs.harvard.edu/abs/2015Ap&SS.357....4K} {357, 4}

\bibitem[\protect\citeauthoryear{{Lauffer}, {Romero}  \& {Kepler}}{{Lauffer} et~al.}{2018}]{2018MNRAS.480.1547L}
{Lauffer} G.~R.,  {Romero} A.~D.,   {Kepler} S.~O.,  2018, \mn@doi [\mnras] {10.1093/mnras/sty1925}, \href {https://ui.adsabs.harvard.edu/abs/2018MNRAS.480.1547L} {480, 1547}

\bibitem[\protect\citeauthoryear{Le}{Le}{2019}]{Le:2019ijj}
Le T.~D.,  2019, \mn@doi [Chin. J. Phys.] {10.1016/j.cjph.2019.10.007}, 62, 252

\bibitem[\protect\citeauthoryear{{Le}}{{Le}}{2021}]{Le2021}
{Le} T.~D.,  2021, \mn@doi [Journal of High Energy Astrophysics] {10.1016/j.jheap.2021.01.001}, \href {https://ui.adsabs.harvard.edu/abs/2021JHEAp..29...43L} {29, 43}

\bibitem[\protect\citeauthoryear{{Lemos}, {Gon{\c{c}}alves}, {Carvalho}  \& {Alcaniz}}{{Lemos} et~al.}{2025}]{2025JCAP...01..059L}
{Lemos} T.,  {Gon{\c{c}}alves} R.,  {Carvalho} J.,   {Alcaniz} J.,  2025, \mn@doi [\jcap] {10.1088/1475-7516/2025/01/059}, \href {https://ui.adsabs.harvard.edu/abs/2025JCAP...01..059L} {2025, 059}

\bibitem[\protect\citeauthoryear{Levshakov, Combes, Boone, Agafonova, Reimers  \& Kozlov}{Levshakov et~al.}{2012}]{Levshakov:2012kv}
Levshakov S.~A.,  Combes F.,  Boone F.,  Agafonova I.~I.,  Reimers D.,   Kozlov M.~G.,  2012, \mn@doi [Astron. Astrophys.] {10.1051/0004-6361/201219042}, 540, L9

\bibitem[\protect\citeauthoryear{{Levshakov}, {Henkel}, {Reimers}  \& {Molaro}}{{Levshakov} et~al.}{2014}]{Levshakov:2013oja}
{Levshakov} S.~A.,  {Henkel} C.,  {Reimers} D.,   {Molaro} P.,  2014, \mn@doi [\memsai] {10.48550/arXiv.1307.8266}, \href {https://ui.adsabs.harvard.edu/abs/2014MmSAI..85...90L} {85, 90}

\bibitem[\protect\citeauthoryear{Levshakov, Ng, Henkel  \& Mookerjea}{Levshakov et~al.}{2017}]{Levshakov:2017ivg}
Levshakov S.~A.,  Ng K.~W.,  Henkel C.,   Mookerjea B.,  2017, \mn@doi [Mon. Not. Roy. Astron. Soc.] {10.1093/mnras/stx1782}, 471, 2143

\bibitem[\protect\citeauthoryear{{Magano}, {Vilas Boas}  \& {Martins}}{{Magano} et~al.}{2017}]{2017PhRvD..96h3012M}
{Magano} D.~M.~N.,  {Vilas Boas} J.~M.~A.,   {Martins} C.~J.~A.~P.,  2017, \mn@doi [\prd] {10.1103/PhysRevD.96.083012}, \href {https://ui.adsabs.harvard.edu/abs/2017PhRvD..96h3012M} {96, 083012}

\bibitem[\protect\citeauthoryear{{Martins}}{{Martins}}{2017}]{2017RPPh...80l6902M}
{Martins} C.~J.~A.~P.,  2017, \mn@doi [Reports on Progress in Physics] {10.1088/1361-6633/aa860e}, \href {https://ui.adsabs.harvard.edu/abs/2017RPPh...80l6902M} {80, 126902}

\bibitem[\protect\citeauthoryear{{Mazzali}, {Chugai}, {Turatto}, {Lucy}, {Danziger}, {Cappellaro}, {della Valle}  \& {Benetti}}{{Mazzali} et~al.}{1997}]{1997MNRAS.284..151M}
{Mazzali} P.~A.,  {Chugai} N.,  {Turatto} M.,  {Lucy} L.~B.,  {Danziger} I.~J.,  {Cappellaro} E.,  {della Valle} M.,   {Benetti} S.,  1997, \mn@doi [\mnras] {10.1093/mnras/284.1.151}, \href {http://adsabs.harvard.edu/abs/1997MNRAS.284..151M} {284, 151}

\bibitem[\protect\citeauthoryear{{Monteiro}, {Ferreira}, {Juli{\~a}o}  \& {Martins}}{{Monteiro} et~al.}{2014}]{2014MmSAI..85..113M}
{Monteiro} A.~M.~R.~V.~L.,  {Ferreira} M.~C.,  {Juli{\~a}o} M.~D.,   {Martins} C.~J.~A.~P.,  2014, \memsai, \href {https://ui.adsabs.harvard.edu/abs/2014MmSAI..85..113M} {85, 113}

\bibitem[\protect\citeauthoryear{{Mosquera} \& {Civitarese}}{{Mosquera} \& {Civitarese}}{2017}]{2017PhRvC..96d5802M}
{Mosquera} M.~E.,  {Civitarese} O.,  2017, \mn@doi [\prc] {10.1103/PhysRevC.96.045802}, \href {https://ui.adsabs.harvard.edu/abs/2017PhRvC..96d5802M} {96, 045802}

\bibitem[\protect\citeauthoryear{Murphy \& Cooksey}{Murphy \& Cooksey}{2017}]{Murphy:2017xaz}
Murphy M.~T.,  Cooksey K.~L.,  2017, \mn@doi [Mon. Not. Roy. Astron. Soc.] {10.1093/mnras/stx1949}, 471, 4930

\bibitem[\protect\citeauthoryear{{Nakashima}, {Ichikawa}, {Nagata}  \& {Yokoyama}}{{Nakashima} et~al.}{2010}]{nakashima2010constraining}
{Nakashima} M.,  {Ichikawa} K.,  {Nagata} R.,   {Yokoyama} J.,  2010, \mn@doi [\jcap] {10.1088/1475-7516/2010/01/030}, \href {https://ui.adsabs.harvard.edu/abs/2010JCAP...01..030N} {2010, 030}

\bibitem[\protect\citeauthoryear{Rahmani, Srianand, Gupta, Petitjean, Noterdaeme  \& Vasquez}{Rahmani et~al.}{2012}]{Rahmani:2012ze}
Rahmani H.,  Srianand R.,  Gupta N.,  Petitjean P.,  Noterdaeme P.,   Vasquez D.~A.,  2012, \mn@doi [Mon. Not. Roy. Astron. Soc.] {10.1111/j.1365-2966.2012.21503.x}, 425, 556

\bibitem[\protect\citeauthoryear{{Rotondo}, {Rueda}, {Ruffini}  \& {Xue}}{{Rotondo} et~al.}{2011a}]{2011PhRvC..83d5805R}
{Rotondo} M.,  {Rueda} J.~A.,  {Ruffini} R.,   {Xue} S.~S.,  2011a, \mn@doi [\prc] {10.1103/PhysRevC.83.045805}, \href {https://ui.adsabs.harvard.edu/abs/2011PhRvC..83d5805R} {83, 045805}

\bibitem[\protect\citeauthoryear{{Rotondo}, {Rueda}, {Ruffini}  \& {Xue}}{{Rotondo} et~al.}{2011b}]{Rotondo:2011zz}
{Rotondo} M.,  {Rueda} J.~A.,  {Ruffini} R.,   {Xue} S.-S.,  2011b, \mn@doi [\prd] {10.1103/PhysRevD.84.084007}, \href {https://ui.adsabs.harvard.edu/abs/2011PhRvD..84h4007R} {84, 084007}

\bibitem[\protect\citeauthoryear{{Salpeter}}{{Salpeter}}{1961}]{1961ApJ...134..669S}
{Salpeter} E.~E.,  1961, \mn@doi [\apj] {10.1086/147194}, \href {https://ui.adsabs.harvard.edu/abs/1961ApJ...134..669S} {134, 669}

\bibitem[\protect\citeauthoryear{{Salpeter} \& {van Horn}}{{Salpeter} \& {van Horn}}{1969}]{1969ApJ...155..183S}
{Salpeter} E.~E.,  {van Horn} H.~M.,  1969, \mn@doi [\apj] {10.1086/149858}, \href {https://ui.adsabs.harvard.edu/abs/1969ApJ...155..183S} {155, 183}

\bibitem[\protect\citeauthoryear{{Sarmah}, {Kalita}  \& {Wojnar}}{{Sarmah} et~al.}{2022}]{2022PhRvD.105b4028S}
{Sarmah} L.,  {Kalita} S.,   {Wojnar} A.,  2022, \mn@doi [\prd] {10.1103/PhysRevD.105.024028}, \href {https://ui.adsabs.harvard.edu/abs/2022PhRvD.105b4028S} {105, 024028}

\bibitem[\protect\citeauthoryear{{Scalzo} et~al.,}{{Scalzo} et~al.}{2010}]{2010ApJ...713.1073S}
{Scalzo} R.~A.,  et~al., 2010, \mn@doi [\apj] {10.1088/0004-637X/713/2/1073}, \href {http://adsabs.harvard.edu/abs/2010ApJ...713.1073S} {713, 1073}

\bibitem[\protect\citeauthoryear{{Shapiro} \& {Teukolsky}}{{Shapiro} \& {Teukolsky}}{1983}]{1983bhwd.book.....S}
{Shapiro} S.~L.,  {Teukolsky} S.~A.,  1983, {Black holes, white dwarfs and neutron stars: The physics of compact objects}.
Wiley-VCH, New York, \mn@doi{10.1002/9783527617661}

\bibitem[\protect\citeauthoryear{Srianand, Gupta, Petitjean, Noterdaeme  \& Ledoux}{Srianand et~al.}{2010}]{Srianand:2010un}
Srianand R.,  Gupta N.,  Petitjean P.,  Noterdaeme P.,   Ledoux C.,  2010, \mn@doi [Mon. Not. Roy. Astron. Soc.] {10.1111/j.1365-2966.2010.16574.x}, 405, 1888

\bibitem[\protect\citeauthoryear{{Toniato}, {Rodrigues}  \& {Wojnar}}{{Toniato} et~al.}{2020}]{2020PhRvD.101f4050T}
{Toniato} J.~D.,  {Rodrigues} D.~C.,   {Wojnar} A.,  2020, \mn@doi [\prd] {10.1103/PhysRevD.101.064050}, \href {https://ui.adsabs.harvard.edu/abs/2020PhRvD.101f4050T} {101, 064050}

\bibitem[\protect\citeauthoryear{Ubachs, Bagdonaite, Salumbides, Murphy  \& Kaper}{Ubachs et~al.}{2016}]{Ubachs:2015fro}
Ubachs W.,  Bagdonaite J.,  Salumbides E.~J.,  Murphy M.~T.,   Kaper L.,  2016, \mn@doi [Rev. Mod. Phys.] {10.1103/RevModPhys.88.021003}, 88, 021003

\bibitem[\protect\citeauthoryear{Uniyal, Kalita  \& Chakrabarti}{Uniyal et~al.}{2023}]{Uniyal:2023bff}
Uniyal A.,  Kalita S.,   Chakrabarti S.,  2023, \mn@doi [Mon. Not. Roy. Astron. Soc.] {10.1093/mnras/stad3123}, 527, 232

\bibitem[\protect\citeauthoryear{{Webb}, {Lee}, {Milakovi{\'c}}, {Dougan}, {Dzuba}  \& {Flambaum}}{{Webb} et~al.}{2024}]{Webb:2024}
{Webb} J.~K.,  {Lee} C.-C.,  {Milakovi{\'c}} D.,  {Dougan} D.,  {Dzuba} V.~A.,   {Flambaum} V.~V.,  2024, \mn@doi [arXiv e-prints] {10.48550/arXiv.2410.00684}, \href {https://ui.adsabs.harvard.edu/abs/2024arXiv241000684W} {p. arXiv:2410.00684}

\bibitem[\protect\citeauthoryear{{Zel'dovich} \& {Novikov}}{{Zel'dovich} \& {Novikov}}{1966}]{YaBZel'dovich_1966}
{Zel'dovich} Y.~B.,  {Novikov} I.~D.,  1966, \mn@doi [Soviet Physics Uspekhi] {10.1070/PU1966v008n04ABEH002990}, \href {https://ui.adsabs.harvard.edu/abs/1966SvPhU...8..522Z} {8, 522}

\bibitem[\protect\citeauthoryear{de Carvalho, Rotondo, Rueda  \& Ruffini}{de~Carvalho et~al.}{2013}]{deCarvalho:2013rea}
de Carvalho S.~M.,  Rotondo M.,  Rueda J.~A.,   Ruffini R.,  2013, \mn@doi [Int. J. Mod. Phys. Conf. Ser.] {10.1103/PhysRevC.89.015801}, 23, 244

\bibitem[\protect\citeauthoryear{{de Carvalho}, {Rotondo}, {Rueda}  \& {Ruffini}}{{de Carvalho} et~al.}{2014}]{de2014relativistic}
{de Carvalho} S.~M.,  {Rotondo} M.,  {Rueda} J.~A.,   {Ruffini} R.,  2014, \mn@doi [\prc] {10.1103/PhysRevC.89.015801}, \href {https://ui.adsabs.harvard.edu/abs/2014PhRvC..89a5801D} {89, 015801}

\makeatother
\end{thebibliography}

%%%%%%%%%%%%%%%%%%%%%%%%%%%%%%%%%%%%%%%%%%%%%%%%%%%%%%%%%%%%%%%%%%%%%%%%%%%%%%%%%%%%%%%%%%%%%%%%%%%

% Don't change these lines
\bsp	% typesetting comment
\label{lastpage}
\end{document}